\documentclass[%
 reprint,
showpacs,preprintnumbers,
 amsmath,amssymb,
 aps,
prl,
]{revtex4-1}

\usepackage{graphicx}
\usepackage{dcolumn}
\usepackage{bm}
\usepackage{ulem}

\begin{document}

\preprint{APS/123-QED}

\title{
Spontaneous Parity Breaking in Spin-Orbital Coupled Systems}

\author{Satoru Hayami,$^1$ Hiroaki Kusunose,$^2$ and Yukitoshi Motome$^1$}
\affiliation{%
 $^1$Department of Applied Physics, University of Tokyo, Tokyo 113-8656, Japan \\
 $^2$Department of Physics, Ehime University, Matsuyama 790-8577, Japan
}%

\begin{abstract}
Effects of spontaneous parity breaking by charge, spin, and orbital orders are investigated in a two-band Hubbard model on a honeycomb lattice. 
This is a minimal model in which the inter-orbital hopping, atomic spin-orbit coupling, and strong electron correlation give rise to fascinating properties, such as the magnetoelectric effects, quantum spin Hall effect, and spin or valley splitting in the band structure. 
We perform the symmetry analysis of possible broken-parity states and the mean-field analysis of their competition. 
We find that the model at 1/4 filling exhibits a spin-orbital composite ordered state and a charge ordered state, in addition to a paramagnetic quantum spin-Hall insulator. 
We show that the composite ordered phase exhibits two types of magnetoelectric responses. 
The charge ordered state shows spin splitting in the band structure, while the topological nature varies depending on electron correlations. 
\end{abstract}
\maketitle

The spin-orbit coupling (SOC) has drawn considerable attention in condensed matter physics since it gives rise to various fascinating phenomena, such as the Dirac electrons at the surface of topological insulators~\cite{Hasan_RevModPhys.82.3045,moore2010birth,qi2011topological}, spin Hall effect~\cite{hirsch1999spin,sinova2004universal,bernevig2006quantum}, and the noncentrosymmetric superconductivity~\cite{Bauer_PhysRevLett.92.027003,Bauer_Sigrist201201}. 
A key concept in these phenomena is peculiar spin-orbital entanglement by the antisymmetric SOC, which originates from the atomic SOC in the absence of spatial inversion (parity) symmetry. 
Such spin-orbital entangled physics has been found in a lot of real materials experimentally. 
Monolayer dichalcogenides $MX_2$ ($M$: transition metal, $X$: chalcogen)~\cite{wilson1969transition,li2007electronic,PhysRevB.79.115409,xiao2012coupled} are one of the recent examples, in which the intriguing spin and valley physics has attracted much interests in the light of applications to electronic devices. 

On the other hand, electron correlations bring a new aspect by stabilizing various spontaneous electronic orders in the spin-orbital coupled systems. 
A representative example is multiferroics, which are magnetic insulators showing magnetoelectric (ME) responses as a consequence of the interplay between charge, spin, and orbital degrees of freedom. 
In the multiferroic compounds, a magnetic ordering breaks spatial inversion symmetry as well as time reversal symmetry, and induces a spontaneous electric polarization~\cite{kimura2003magnetic,fiebig2005revival,cheong2007multiferroics,khomskii2009classifying}. 

\begin{figure}[t]
\centering
\includegraphics[width=1.0 \hsize]{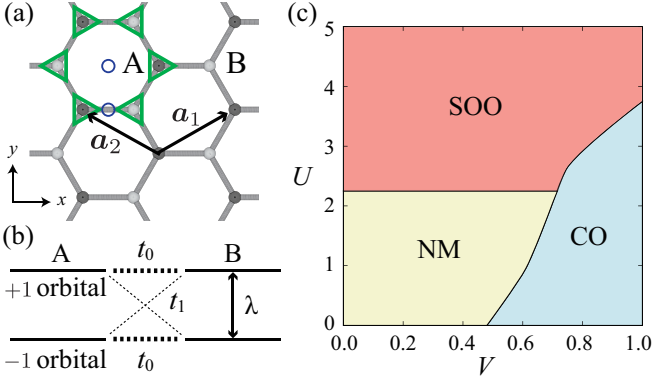} 
\caption{ 
\label{Fig:souzu}
(Color online) 
(a) Schematic picture of a honeycomb lattice; 
the primitive translation vectors are $\bm{a}_1 = (\sqrt{3}/2, 1/2)$ and $\bm{a}_2 = (-\sqrt{3}/2, 1/2)$. 
Open circles (triangles) indicate the inversion centers (the parity-breaking sites). 
(b) Schematic picture of the energy levels of the two-band model Hamiltonian in Eq.~(\ref{Eq:H0}). 
(c) Ground-state phase diagram of the model in Eqs.~(\ref{Eq:H0}) and~(\ref{eq:H1}) by the mean-field calculations. 
We take $t_0=t_1=\lambda=0.5$ and $J_{\rm H}=0.1U$. 
NM, CO, and SOO stand for the nonmagnetic, 
charge ordered, and spin-orbital composite ordered states, respectively. 
}
\end{figure}

In this Rapid Communication, we propose yet another interesting situation that arises in 
an interplay between electron correlations and proper lattice structures, namely, 
systems where the inversion symmetry is preserved globally but broken intrinsically at atomic sites. 
We refer to this as {\it local} parity breaking. 
For instance, a honeycomb lattice possesses the inversion symmetry with respect to the bond centers and the hexagon centers, but breaks it at the lattice sites, as shown in Fig.~\ref{Fig:souzu}(a). 
In such systems, the antisymmetric SOC is hidden at each site in a site-dependent form. 
Once a long-range order occurs in charge, spin, and orbital degrees of freedom of electrons in a way of breaking the global inversion symmetry, a net antisymmetric SOC emerges. 
Such an emergent interaction can bring about new ME effects and transport phenomena spontaneously that have not been seen in the ordinary multiferroic insulators. 

We present a theoretical analysis of symmetry broken states in a microscopic model with the local parity breaking. 
Particularly, we consider a minimal two-band model on a honeycomb lattice, in which the global inversion symmetry can be broken by spontaneous bipartite electronic orders. 
First, by the symmetry analysis, we categorize possible charge, spin, and orbital staggered orders into seven different classes. 
For these classes, we investigate the influence of symmetry breaking, focusing on the ME effects. 
Next, we study the ground state and finite-temperature properties by the mean-field approximation. 
As a result, we find that the system at 1/4 electron filling exhibits a peculiar spin-orbital composite ordered state and a charge ordered state, in addition to a paramagnetic state. 
The composite ordered state shows two types of ME responses to an electric field. 
Meanwhile, the paramagnetic state and the charge-ordered state exhibit the quantum spin Hall effect. 
The latter also shows antisymmetric spin splitting in the band structure.

Let us begin with a minimal model Hamiltonian on the honeycomb lattice [Fig.~\ref{Fig:souzu}(a)] for describing the spontaneous parity breaking. 
We include the orbital degree of freedom for $d$-electron systems under a crystalline electric field, such as, e.g., the trigonal, trigonal prismatic, and square antiprismatic configuration of the ligands. 
Specifically, in the trigonal prismatic case, the atomic energy levels are split into the lowest-energy levels with $m=\pm 2$ ($d_{x^2-y^2}$ and $d_{xy}$ orbitals), the middle one with $m=0$ ($d_z^2$), and the highest ones with $m=\pm 1$ ($d_{zx}$ and $d_{yz}$).
Note that this level scheme is realized in $MX_2$~\cite{xiao2012coupled}, while the lattice of $M$ cations is a stacked triangular lattice. 

Assuming that all the other energy scales are smaller than the level splitting, we consider only two relevant orbitals $m=\pm1$ for simplicity. 
A generalization for $m=\pm2$ case is straightforward. 
Then, the one-body part of our model is given by 
\begin{eqnarray}
\label{Eq:H0}
\mathcal{H}_{0} 
&=& -t_0 \sum_{\bm{k}}\sum_{m} \sum_{\sigma} 
(  \gamma_{0\bm{k}} c_{{\rm A}\bm{k}
m\sigma}^{\dagger} c_{{\rm B}\bm{k} 
m\sigma} + {\rm H.c.}) \nonumber 
\\
& &-t_1 \sum_{\bm{k}}\sum_{m} \sum_{\sigma}  
(\gamma_{m\bm{k}}  c_{{\rm A}\bm{k}
m\sigma}^{\dagger} c_{{\rm B}\bm{k} - 
m\sigma} + {\rm H.c.}) \nonumber
\\ 
& & +\frac{\lambda}{2}\sum_{s} 
\sum_{\bm{k}}\sum_{m} \sum_{\sigma}
c_{s \bm{k} 
m\sigma}^{\dagger} (m\sigma) 
c_{s \bm{k} 
m\sigma}, 
\end{eqnarray}
where $c_{s \bm{k} m\sigma}^{\dagger}$ ($c_{s \bm{k} m\sigma}$) is the creation (annihilation) operator for sublattice $s=$ A 
or B, respectively, 
wave number $\bm{k}$, orbital $m=\pm 1$, and spin $\sigma = \uparrow$ or $\downarrow$. 
The first and second terms represent the intra- and inter-orbital 
hoppings between nearest-neighbor sites, respectively [see Fig.~\ref{Fig:souzu}(b)]. 
The third term in Eq.~(\ref{Eq:H0}) represents the atomic SOC; as we consider only $m=\pm 1$, 
this term has a nonzero matrix element for the diagonal $z$ component in terms of orbitals. 

The $\bm{k}$ dependence 
in the hopping terms is given by 
\begin{align} 
\gamma_{ n\bm{k}} &= e^{i \bm{k}\cdot \bm{\eta}_1 } +  \omega^{ 
-2 n} e^{{\rm i} \bm{k}\cdot \bm{\eta}_2 }+ \omega^{2n} 
e^{{\rm i} \bm{k}\cdot \bm{\eta}_3 }
=\gamma_{-n,-\bm{k}}^{*}, 
\label{eq:gamma}
\end{align}
where $\omega = e^{2\pi {\rm i}/3}$;
$\bm{\eta}_1 = (\bm{a}_1 - \bm{a}_2)/3$, $\bm{\eta}_2 = (\bm{a}_1 + 2 \bm{a}_2)/3$, and $\bm{\eta}_3 =-(2 \bm{a}_1 + \bm{a}_2)/3$ [$\bm{a}_1$ and $\bm{a}_2$ are primitive translational vectors as shown in Fig.~\ref{Fig:souzu}(a)]. 
The additional phase factors in Eq.~(\ref{eq:gamma}) come from the angular-momentum transfers 
between orbitals. Note that $\gamma_{n\bm{k}}^{}$ play an important role, i.e., 
their combination arises antisymmetric $\bm{k}$ dependences which give rise to the spin or valley splitting once a broken-parity electronic order takes place in the presence of the SOC. 

The Hamiltonian $\mathcal{H}_0$ in Eq.~(\ref{Eq:H0}) has five symmetries in addition to the lattice translational symmetry: 
spatial inversion ($\mathcal{P}$), time reversal ($\mathcal{T}$), real-space $2\pi/3$ rotation around the $z$ axis ($\mathcal{R}$), mirror for the $xz$ plane ($\mathcal{M}$), and mirror symmetry for the $yz$ plane~\cite{comment_yz}. 
Each symmetry operation is represented by a combination of three Pauli matrices: 
$\bm{\rho}$ for sublattice, $\bm{\sigma}$ for spin, $\bm{\tau}$ for orbital indices, as follows. 
The spatial inversion $\mathcal{P}$ is represented by $\rho_x$. 
The antiunitary time-reversal operation $\mathcal{T}$ is represented by ${\rm i} \sigma_{y}\tau_x 
K$ ($K$ is a complex conjugate operator). 
The threefold rotation $\mathcal{R}$ is represented by $e^{2 \pi {\rm i} \tau_z /3}$ 
with the cyclic permutation of the site indices. 
The mirror operation $\mathcal{M}$ is represented by ${\rm i} \sigma_{z}$, which is obtained by a product of operations: $k_y$ inversion ($k_y \rightarrow -k_y$ by $\sigma_{x}\tau_{x}$), orbital inversion ($\pm 1 \rightarrow \mp 1$ by $\tau_{x}$), and spin inversion ($\{\sigma_x,\sigma_{z}\} \rightarrow -\{\sigma_x,\sigma_{z}\}$ by $\sigma_{y}$). 

\begin{table}[t]
  \begin{tabular}{|c|c|c|c|c|c||c|c|}
 \hline
\#& O.P. & $\mathcal{P}$ & $\mathcal{T}$& $\mathcal{R}$& $\mathcal{M}$  & ME(u) & ME(s) \\
\hline
\hline
1 & CO, $zz$-SOO & $\times$ & $\bigcirc$ & $\bigcirc$ & $\bigcirc$ & --- & --- \\ \hline
2 & $x/y$-OO & $\times$ & $\bigcirc$ & $\times$ & $\bigcirc$ & --- & $\checkmark$ \\ \hline
3 & $xz/yz$-SOO & $\times$ & $\bigcirc$ & $\bigcirc$ & $\times$ &  --- & --- \\ \hline
4 & $z$-SO, $z$-OO & $\times$ & $\times$ & $\bigcirc$ & $\bigcirc$ & --- & --- \\ \hline
5 & $zx/zy$-SOO & $\times$ & $\times$ & $\times$ & $\bigcirc$ & --- & $\checkmark$ \\ \hline
6 & $x/y$-SO & $\times$ & $\times$ & $\bigcirc$ & $\times$ & --- & --- \\ \hline
7 & $xx/yy/xy/yx$-SOO & $\times$ & $\times$ & $\times$ & $\times$ & $\checkmark$ & $\checkmark$ \\ \hline
\end{tabular}
\caption{
Seven symmetry classes of 16 staggered order parameters categorized in terms of the presence ($\bigcirc$) or absence ($\times$) of the four 
symmetries of the system: spatial inversion ($\mathcal{P}$), time reversal ($\mathcal{T}$), $2\pi/3$ rotation around the $z$ axis ($\mathcal{R}$), and mirror symmetry for the $xz$ plane ($\mathcal{M}$).
CO, SO, OO, and SOO represent charge, spin, orbital, and spin-orbital orders, respectively, and the component prefixes denote the type of orders; see the text for details. 
In the columns for ME(u) and ME(s), the checkmark ($\checkmark$) shows the nonzero uniform and staggered ME effects, respectively. 
SOO in Fig.~\ref{Fig:souzu}(c) corresponds to \#7. 
}
\label{tab}
\end{table}

With these preliminaries, we examine possible electronic orders from the symmetry point of view. 
We here consider only the staggered electronic orders on the honeycomb lattice, which accompany a global parity breaking;
these orders are described in terms of $\rho_{z}$ with the ordering wave number $\bm{Q}=0$. 
There are sixteen candidates for such staggered orders $\Lambda^{\alpha}_{\;\beta}=\sigma_{\alpha}\tau_{\beta}$ ($\alpha,\beta=0,x,y,z$): a charge order $\Lambda^{0}_{\;0}$ (CO), three spin orders $\Lambda^{\mu}_{\;0}$ ($\mu$-SO), three orbital orders $\Lambda^{0}_{\;\nu}$ ($\nu$-OO), and nine spin-orbital composite orders $\Lambda^{\mu}_{\;\nu}$ ($\mu\nu$-SOO). 
Here, $\sigma_{0}$, $\tau_0$, and $ \rho_0$ are $2\times 2$ unit matrices. 
Note that $\tau_{x}$ and $\tau_{y}$ correspond to the electric quadrupole operators $l_{x}^{2}-l_{y}^{2}$ and $l_{x}l_{y}+l_{y}l_{x}$, respectively, while $\tau_{z}$ the (magnetic) orbital angular-momentum operator $l_{z}$ in the $m=\pm1$ subspace. 
The Hamiltonian for symmetry-breaking field corresponding to these orders is given in a general form as 
\begin{align}
\label{eq: Ham_MF}
\tilde{\mathcal{H}}_{1}  
=   
-h \sum_{s}\sum_{\bm{k}} 
\sum_{m m', \sigma \sigma'} c^{\dagger}_{s  \bm{k} m\sigma} 
\left[p(s) 
\Lambda^{\alpha}_{\;\beta}
\right]^{\sigma\sigma'}_{mm'}
c_{s \bm{k} m'\sigma'}, 
\end{align}
where $h$ is the magnitude of the symmetry-breaking field and $p(s)=+1$ $(-1)$ for $s=$A (B). 

These staggered orders break the symmetries $\mathcal{T}$, $\mathcal{R}$, and $\mathcal{M}$ in a different way. 
We categorize them into seven classes with respect to the four symmetries, as summarized in Table~\ref{tab}. 
This symmetry analysis will provide a useful reference for macroscopic physical properties, such as ME effects. 

As a representative example, we show two different types of ME effects in Table~\ref{tab}: 
the uniform and staggered responses to an electric field. 
Here, we complete the table by computing the ME effects for each ordered state on the basis of the linear response theory [see Eq.~(\ref{Eq:Mangetoelectric effect})]. 
The staggered ME effect was discussed for the systems with the local parity breaking in which the antisymmetric SOC exists irrespective of the electronic state~\cite{Yanase:JPSJ.83.014703,Hayami_toroidal}. 
In the present case, however, the antisymmetric SOC does not contribute to the linear ME term as long as the rotational symmetry $\mathcal{R}$ is preserved.
Once breaking it, an additional antisymmetric SOC appears in the linear ME term giving rise to the staggered ME response as shown in Table~\ref{tab}. 
On the other hand, for the uniform ME effect, the time reversal symmetry $\mathcal{T}$ must be broken at least. 
Furthermore, the breaking of both $\mathcal{R}$ and $\mathcal{M}$ is necessary for the honeycomb lattice. 
Remarkably, among 16 possible electronic orders, only the \#7 SOO states exhibit both the uniform and staggered ME effects.

Next, we examine whether such SOOs are stabilized 
in our minimal model by including electron correlations. 
We introduce Coulomb interactions to the Hamiltonian in Eq.~(\ref{Eq:H0}), whose form is given in the real-space representation: 
\begin{align}
\mathcal{H}_{1}&=\sum_{i}\sum_{m n m' n'} \sum_{\sigma \sigma'}
\frac{U_{m n m' n'}}{2} c_{ i m\sigma}^{\dagger}c_{ i n\sigma'}^{\dagger}
c_{ i n'\sigma'} c_{ i m' \sigma}  \nonumber \\
& + \sum_{\langle i, j \rangle}\sum_{mm'} \sum_{\sigma \sigma'} 
V n_{ i m \sigma}n_{j m' \sigma'}, 
\label{eq:H1}
\end{align}
where $n_{im\sigma} = c_{im\sigma}^\dagger c_{im\sigma}$. 
The first term stands for the on-site Coulomb interaction; 
we take $U_{mmmm}=U$, and $U_{mnmn}= U-2J_{{\rm H}}$, $U_{mnnm} =U_{mmnn}=J_{{\rm H}}$ ($m\neq n$), where $U$ is the on-site repulsion and $J_{{\rm H}}$ is the Hund's-rule coupling, respectively. 
The second term is the Coulomb interaction between nearest-neighbor sites, which is introduced to stabilize CO as a competing order.

We study the model $\mathcal{H}_0+\mathcal{H}_1$ by the mean-field approximation. 
We adopt the Hartree-Fock approximation to decouple the on-site Coulomb interaction, while the nearest-neighbor Coulomb repulsion is treated by the Hartree approximation. 
We employ two-site unit cell, and calculate the mean fields by taking the sum over $64 \times 64$ grid points in the first Brillouin zone. 
Hereafter, we take $t_0 = t_1 = \lambda =0.5$ and $J_{{\rm H}}=0.1 U$. 
We have explored the possibility of bipartite orders in Table I for the half-filling and 1/4-filling cases. 
Below, we focus on the latter case (one electron per site on average), as it includes the interesting \#7 SOO phase. 

\begin{figure}[t]
\centering
\includegraphics[width=1.0 \hsize]{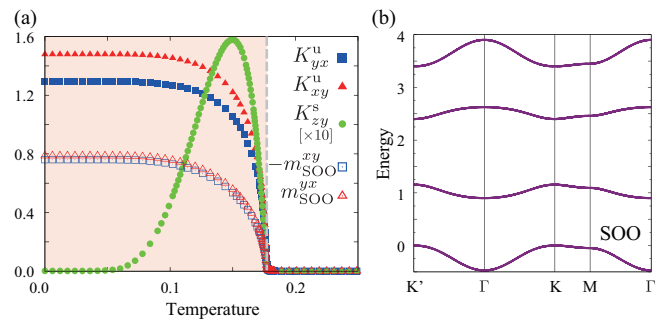} 
\caption{
(Color online) 
(a) Temperature dependence of the ME coefficients, $K^{\rm u}_{yx}$, $K^{\rm u}_{xy}$, and $K^{\rm s}_{zy}$, and 
the order parameters, $m^{xy}_{{\rm SOO}}$ and $m^{yx}_{{\rm SOO}}$ in the SOO region in Fig.~\ref{Fig:souzu}(c). 
The data are calculated at $U=2.5$ and $V=0$. 
The vertical dashed line shows the transition temperature. 
(b) Electronic band structure for the same parameters at $T=0$. 
The results are shown along the high-symmetry lines in the Brillouin zone [see the inset of Fig.~\ref{Fig:band}(b)]. 
The Fermi level is set to zero. 
\label{Fig:Temp}
}
\end{figure}

Figure~\ref{Fig:souzu}(c) shows the ground-state phase diagram at 1/4 filling. 
The phase diagram includes three insulating phases: the nonmagnetic (NM) phase in the small $U$ and $V$ region, CO phase in the large $V$ region, 
and SOO phase in the large $U$ region. 
The last one is the \#7 SOO that exhibits both the uniform and staggered ME responses, as shown in Table~\ref{tab}. 
Indeed, both ME responses become nonzero associated with SOO, as shown in Fig.~\ref{Fig:Temp}(a). 
The result shows the temperature dependence of the ME coefficients obtained by the mean-field approximation at $U=2.5$ and $V=0$. 
Here, we compute the linear response function in the form 
\begin{align}
K^{\rm T}_{\mu\nu} 
=\frac{
2\pi}{{\rm i} V_{0}} \sum_{\alpha\beta \bm{k}} \frac{f(\epsilon_{
\beta\bm{k}})-f(\epsilon_{\alpha\bm{k}})}{\epsilon_{\beta\bm{k}}-\epsilon_{ \alpha\bm{k}}} 
\frac{m_{{\rm T}\mu,\bm{k}}^{\beta\alpha} J_{\nu,\bm{k}}^{\alpha\beta} 
}{\epsilon_{\beta\bm{k}}-\epsilon_{\alpha\bm{k}}+{\rm i} \delta}, 
\label{Eq:Mangetoelectric effect}
\end{align}
where $V_{0}$ the system volume, $f(\epsilon)$ is the Fermi distribution function, and
$\epsilon_{\alpha\bm{k}}$ and $|\alpha\bm{k} \rangle$ are the eigenvalue and eigenstate of ${\cal H}_0+{\cal H}_1$ at the mean-field level. 
Here, $m_{{\rm T}\mu,\bm{k}}^{\beta\alpha}=\langle \beta\bm{k} | \rho_{\rm T}\Lambda^{\mu}_{\;0}  | \alpha\bm{k} \rangle$ 
[T=u ($\rho_0$) or s ($\rho_z$)], and $J_{\nu,\bm{k}}^{\alpha\beta} = \langle \alpha\bm{k} | J_\nu | \beta\bm{k} \rangle$ is the matrix element of 
the current operator. 
Thus, $K^{\rm u}_{\mu\nu}$ ($K^{\rm s}_{\mu\nu}$) represents the coefficient for the uniform (staggered) spin moment along the $\mu$ direction induced by the electric field in the $\nu$ direction. 
In Eq.~(\ref{Eq:Mangetoelectric effect}), we set $(g \mu_{{\rm B}}/2)e/h=1$ ($g$ is the $g$-factor, $\mu_{{\rm B}}$ the Bohr magneton, $e$  the elementary charge, and $h$ the Planck constant) and the damping factor $\delta=0.01$. 

As shown in Fig.~\ref{Fig:Temp}(a), 
$K^{\rm s}_{zy}$ is induced below the critical temperature $T_c \simeq 0.17$ (other staggered components are all zero). 
It shows a peak slightly below $T_c$ and decreases rapidly with 
decreasing temperature. 
This behavior is qualitatively different from the toroidal response in the system with the staggered antisymmetric SOC studied by the authors recently~\cite{Hayami_toroidal}. 
While the toroidal response is already nonzero above $T_c$ and has a peak at $T_c$, the staggered response in the present model becomes nonzero only below $T_c$. 
This is because the nonzero {\it linear-response} $K^{\rm s}_{zy}$ appears only with the lack of the rotational symmetry $\mathcal{R}$, as in the $xy$-SOO state. 
Indeed, the temperature dependence of $K^{\rm s}_{zy}$ is roughly understood by the convolution of the previous toroidal response and that of the order parameter. 

The uniform response, $K^{\rm u}_{\mu\nu}$ ($\mu,\nu=x$ or $y$), also becomes nonzero below $T_c$, as shown in Fig.~\ref{Fig:Temp}(a) (other uniform components are all zero). 
In contrast to $K^{\rm s}_{zy}$, it behaves like the order parameter; in fact, $K^{\rm u}_{yx}$ and $K^{\rm u}_{xy}$ become larger for larger $m^{xy}_{\rm SOO}$ and $m^{yx}_{\rm SOO}$, respectively ($m_{{\rm SOO}}^{\mu\nu}$ is the order parameter for $\mu\nu$-SOO). 
The small difference between them depends on the linear-combination ratio between 
$m^{xy}_{\rm SOO}$ and $m^{yx}_{\rm SOO}$ [see Fig.~\ref{Fig:Temp}(a)]~\cite{comment_OP}.

\begin{figure}[t]
\centering
\includegraphics[width=1.0 \hsize]{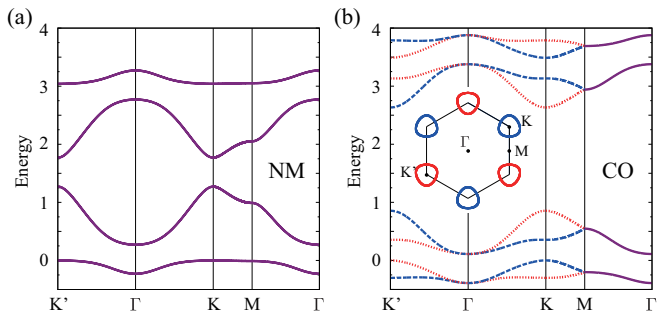} 
\caption{
\label{Fig:band}
(Color online) 
Electronic band structures: (a) for the NM state at $U=V=0 $ and (b) for the CO state at $U=0$ and $V=0.6$. 
In (b), the dashed blue (dotted red) lines show the bands with the up- (down-) spin polarization.
The inset of (b) shows the energy contour below the Fermi level by $0.05$. 
}
\end{figure}

Now let us discuss the other two phases, NM and CO in Fig.~\ref{Fig:souzu}(c). 
They also exhibit interesting transport properties. 
First, we discuss the NM phase. 
Figure~\ref{Fig:band}(a) shows the band structure in this phase; there are four bands, each of which 
is doubly degenerate owing to the presence of both spatial inversion and time reversal symmetries. 
The insulating gap in the NM state derives its origin from both the atomic SOC and inter-orbital hopping. 
Although the band structure is qualitatively similar to that in the SOO state shown in Fig.~\ref{Fig:Temp}(b), the transport property is very different; 
in the NM phase, the hidden antisymmetric SOC gives rise to the quantum spin Hall effect. 
The origin of this quantum spin Hall insulator is essentially the same as that in the single-band Hubbard model with imaginary hopping between next nearest-neighbor sites~\cite{Haldane_PhysRevLett.61.2015,PhysRevLett.95.226801}. 
In the present case, the spin Hall conductivity is quantized at 2 in units of $e /2\pi$. 
We note that the quantized value takes $0$, $2$, and $-4$ depending on $t_1$, $\lambda$, and electron filling. 

With increasing $V$, the NM state changes into the CO state. 
The transition is continuous for $U \lesssim 1.6$. 
Figure~\ref{Fig:band}(b) shows the band structure for the CO state; the dashed (dotted) lines denote the up- (down-) spin component, demonstrating that the antisymmetric spin splitting occurs in the CO state. 
To show this explicitly, we present the energy contour slightly below the top of the highest occupied band in the inset of Fig.~\ref{Fig:band}(b). 
The electronic structure around the K and K$'$ points is similar to that found in the triangular-lattice monolayer dichalcogenides $MX_2$~\cite{xiao2012coupled}. 
In other words, this peculiar electronic state could be created by spontaneous electronic orders for systems with the local parity breaking. 
While this CO insulator for $U \lesssim 1.6$ and smaller $V$ also exhibits the same quantization of the spin Hall conductivity as the NM state, the increase of $V$ leads to a gap closing at the K and K$'$ points, above which the spin Hall conductivity becomes zero. 
These properties suggest a potential topological switching by tuning a magnitude of the order parameter via a change of temperature or coupling constants. 
The detailed analysis of the quantum spin Hall effect in the CO-NM phases will be reported elsewhere. 

Finally, let us remark on other peculiar orders. 
As pointed out in Ref.~\cite{li2013coupling}, the staggered spin order 
(the \#4 $z$-SO in Table~\ref{tab}) leads to the valley splitting in the band structure, i.e., the different gap magnitudes at the K and K$'$ points. 
In addition, we find that the \#5 $zx/zy$-SOO in Table~\ref{tab} shows an antisymmetric band deformation with a shift of the band bottom along the K$'$-$\Gamma$-K line. 
Within the present model calculations, however, we could not find a parameter region for stabilizing these orders.

To summarize, we have investigated the effect and stability of electronic orders which break spontaneously the parity symmetry on the basis of a minimal two-band model on the honeycomb lattice. 
Conducting the symmetry analysis of possible staggered orders, we have discussed the emergence of magnetoelectric responses from the symmetry point of view. 
We have also studied the ground state and finite temperature properties of the minimal model by the mean-field approximation. 
We have found the three insulating phases at 1/4 filling: the spin-orbital composite ordered state, the nonmagnetic state, and the charge ordered state, which exhibit two different types of magnetoelectric effects, the quantum spin Hall effect, and antisymmetric spin splitting in the band structure, respectively. 
Our present analysis will provide useful reference for understanding the physics in the systems with the local parity breaking,
which are promising playgrounds showing fascinating electromagnetic and transport properties with peculiar electronic structures by spontaneous electronic orders.

The authors thank T. Arima and H. Tsunetsugu for fruitful discussions. 
SH is supported by Grant-in-Aid for JSPS Fellows. 
This work was supported by Grants-in-Aid for Scientific Research (No. 24340076), the Strategic Programs for Innovative Research (SPIRE), MEXT, and the Computational Materials Science Initiative (CMSI), Japan.

\bibliographystyle{apsrev}
\bibliography{ref}

\begin{thebibliography}{23}
\bibitem[{com({\natexlab{a}})}]{comment_yz}
\bibinfo{note}{We ignore the mirror operation for the $yz$ plane because it is
  always broken for all the staggered orders considered in the present work.}
\bibitem[{com({\natexlab{b}})}]{comment_OP}
\bibinfo{note}{The order parameter in the \#7 SOO states is generally described
  by an arbitrary linear combination of two combined order parameters,
  ($m_{{\rm SOO}}^{xx}$, $m_{{\rm SOO}}^{yy}$) and ($m_{{\rm SOO}}^{xy}$,
  $m_{{\rm SOO}}^{yx}$), where the ratio within each combined order parameter
  is fixed. We here adopt a simple case written by a single component $(m_{{\rm
  SOO}}^{xy}$, $m_{{\rm SOO}}^{yx}$). For the $(m_{{\rm SOO}}^{xx}$, $m_{{\rm
  SOO}}^{yy}$) component, the longitudinal responses $K_{xx}^{u}$ and
  $K_{yy}^{u}$ become nonzero instead of the transverse ones.}
\end{thebibliography}

\end{document}